\documentclass{ws-p9-75x6-50}
\usepackage{amsfonts,amssymb,epsfig}
\usepackage{amsmath}
\def\sqr#1#2{{\vcenter{\hrule height.#2pt\hbox{\vrule width.#2pt
height#1pt \kern#1pt \vrule width.#2pt}\hrule height.#2pt}}}

\def\hook{\hbox{\vrule height0pt width4pt depth0.3pt
\vrule height7pt width0.3pt depth0.3pt \vrule height0pt width2pt
depth0pt} }
\def\br{\begin{eqnarray}}
\def\er{\end{eqnarray}}
\def\brn{\begin{eqnarray*}}
\def\ern{\end{eqnarray*}}
\def\er{\end{eqnarray}}
\def\beq{\begin{equation}}
\def\eeq{\end{equation}}

\def\vt{\vartheta}

\def\L{{\cal{L}}}

\def\a{\alpha}

\def\L{\mathcal{L}}

\title{\bf The generalized teleparallel structure.}
\author{Yakov Itin }
\address{Institute of Mathematics,  Hebrew University of Jerusalem, \\
Givat Ram, Jerusalem 91904, Israel \\E-mail: itin@math.huji.ac.il}

\begin{document}  
\date{}
\maketitle

The teleparallel geometrical structures has evoked in resent time 
a considerable  interest  for various reasons. 
It is considered as a possible physical relevant geometry itself 
as well as an essential part of generalized non-Riemannian theories 
such as the metric - affine gravity. 
Another important subject is the various applications of the frame technique 
in physical theories based on classical (pseudo) Riemannian geometry. \\
Teleparallel structure exists on 
a manifold of a vanishing second Stiefel-Whitney class. 
This is a restrictive condition relative to existence of the Lorentzian 
metrics. It guarantees: \\
1) orientability, time- and space-orientability, \\
2) existence of a unique Riemannian and a unique Lorentzian structures,\\
3) existence of a spinorial structure,\\
4) good posing of the Cauchy problem.\\
Thus the  restriction of the space-time topology is good motivated 
by physical requirements. \\
The teleparallel structures actually used in gravity  are described by an {\it equivalence class} $[\vt^a]$ of coframe fields on a differential manifold. 
The  {\it equivalence relation} on this 
class is constructed by a certain group of transformations. 
The group $G=GL(n,\mathbb R)$ provides the maximal equivalence class: 
it includes all the possible coframes at a point. 
 Certainly, a sufficient large subgroup of $GL(n,\mathbb R)$ can 
also be chosen 
for the equivalence relation. 
Thus an important question for the teleparallel construction is:\\
{\it What subgroups of   the general linear group
 produce relevant physical models?}\\
The choice of  a subgroup of $GL(n,\mathbb R)$ can be described by 
declaration of a set of invariant conditions on a matrix ${G^a}_b$. 
These invariants are  directly  connected with geometrical and physical 
objects.\\
The choice $G=SL(n,\mathbb R)$ provides the maximal equivalence class which 
preserves the volume element structure on $M$.\\
Similarly, the choice $G=SO(n,\mathbb R)$ provides the maximal equivalence 
class which preserves the Riemannian metric structure on $M$.\\
As for the Lorentzian structure it is preserved by the action of the group 
 $G=SO(1,n-1,\mathbb R)$ or of a  suitable subgroup of it. \\
All the structures above are naturally constructed by coframes. \\
The symplectic structure can also be constructed via the coframe field.  
\begin{equation}
w=w_{ab}\vt^a\wedge\vt^b,\qquad w_{ab}=\begin{pmatrix}
0 &  I\\
-I &  0
\end{pmatrix}.
 \end{equation}
This structure is preserved by $Sp(2n,\mathbb R)$ transformations of the coframe field. 
The geometry of a symplectic manifold can be developed in the form 
similar to the Riemannian geometry. 
It means that a torsion free 
connection preserving the symplectic metric can be introduced. 
This is a symplectic analog of the Levi-Civita connection.\\
It is not sufficient to have a symplectic structure on the physical manifold 
because the lengths of the vectors have also to be calculated. 
Thus it is interesting to consider a mixed structure with two different metrics: 
one pseudo (Riemannian) and the second symplectic. Note that such structure  
is in a some sense equivalent  to the teleparallel structure. 
Indeed, the coframe in a general coordinate chart has $n^2$ independent components. 
As for a pseudo(Riemannian) metric it incorporates $\frac {n(n+1)}2$ independent 
products of the coframe components.  The symplectic metric furnishes the 
rest,  
$\frac {n(n-1)}2$ independent products.
Thus  a coframe structure can be considered as a combination of a  
metric $\{g\}$ and a symplectic form $\{w\}$. 
Since $\{g\}$ is reserved for gravity it is plausible to define the 
electromagnetic field by a symplectic form. \\
In order to preserve the (pseudo)Riemannian and the symplectic metrics under 
global transformations of the coframes one has  to consider the intersection 
of the corresponding groups. \\
In the case of a manifold with Euclidean signature the resulting group is well known
$$O(n,\mathbb R) \cap Sp(n,\mathbb R)=U(\frac n 2).$$
For a 4D manifold it turns to be the basic group of the electro-week 
interaction  
$$O(4,\mathbb R) \cap Sp(4,\mathbb R)=U(1)\times SU(2).$$
In the case of a manifold with Lorentzian signature the resulting group 
 $$O(1,n-1,\mathbb R) \cap Sp(n,\mathbb R)$$
involves the boosts. It is non-compact, consequently, it is interesting to look for  a maximal compact sub-group of it. \\
By introducing of a 3-indexed of the torsion ${C^a}_{bc}=e_c\hook e_b\hook d\vt^a$ the general (pseudo)Riemannian quadratic Lagrangian is represented. 
This Lagrangian is a linear combination of 3 independent pieces. 
\begin{equation}\label{RT}
\L_{RT}=\Big(\a_1C_{abc}C^{abc}+\a_2C_{abc}C^{bac}+\a_3C_{a}C^{a}\Big)*1
\end{equation}
It   coincides with the translation invariant  Lagrangian of Rumpf. \\
The quadratic teleparallel Lagrangian for a teleparallel manifold with a symplectic structure is unique 
$$\L_{symp}=w^{nk}{C^m}_{ma}{C^a}_{nk}*1.$$
As for a quadratic Riemannian-symplectic  teleparallel Lagrangian it 
is a linear combination of 16 independent 4-forms, which are constructed explicitly. \\
The field equations for the structures described above will be studied. 
\end{document}